\documentclass[11pt,a4paper]{article}
\usepackage{amstex}
\usepackage{amscd}
\newcommand{\MBB}[1]{{\mathbb{#1}}}
\newcommand{\MCA}[1]{{\mathcal{#1}}}
\newcommand{\MRM}[1]{{\mathrm{#1}}}
\newcommand{\REF}[1]{{(\ref{#1})}}
\begin{document}
\begin{titlepage}
$\phantom{X}$\vspace{2cm}
\begin{center}
{\Large 
Massless picture, massive picture, and symmetry\\ 
in the Gaussian renormalization group
\\[10mm]
}
\end{center}
\begin{center}
{\Large C. Wieczerkowski} \\[10mm]
\end{center}
\begin{center}
Institut f\"ur Theoretische Physik I,
Universit\"at M\"unster, \\
Wilhelm-Klemm-Stra\ss e 9, D-48149 M\"unster, \\
wieczer$@$uni-muenster.de \\[2mm]
\end{center}
\vspace{-10cm}
\hfill MS-TP1-98-02
\vspace{11cm}
\begin{abstract}\noindent
We consider renormalization groups of transformations composed of a 
Gaussian convolution and a field dilatation. As an example, we consider 
perturbations of a single component real Euclidean free field $\phi$ 
with covariance $(-\bigtriangleup)^{-1+\frac{\epsilon}{2}}$. We show
that the renormalization group admits two equivalent formulations called 
massless picture and massive picture respectively. We then show in the 
massive picture that the renormalization group has a symmetry. The symmetry 
consists of global scale transformations composed with certain Gaussian 
convolutions. We translate the symmetry back to the massless picture. The 
relation between the symmetry and the notion of an anomalous dimension is 
briefly discussed.
\end{abstract}
\end{titlepage}

\section{Introduction}
We consider renormalization groups \cite{Wilson/Kogut:1974}
of transformations, which are composed of a Gaussian 
convolution and a field dilatation. Transformations of this 
kind \cite{Gallavotti:1985} are tailor made to study 
perturbations of free fields. Specifically, we consider 
effective field theories given by measures
\begin{equation}
\MRM{d}\rho(\phi)=
   \frac{\MRM{d}\mu_{C_{\infty}}(\phi)\;Z(\phi)}
   {\int\MRM{d}\mu_{C_{\infty}}(\phi)\;Z(\phi)}
\label{0.1}
\end{equation}
on field space $\{\phi:\MBB{R}^{D}\rightarrow\MBB{R}\}$,
where $\MRM{d}\mu_{C_{\infty}}(\phi)$ is the Gaussian measure
with covariance $C_{\infty}$, and where $Z(\phi)$ is a 
perturbation, which is normalized such that $Z(0)=1$. 
We consider perturbations of a massless free field with 
$C_{\infty}(x,y)$ of the order $O\left(\vert x-y\vert^{2-D-\epsilon}
\right)$ at $\vert x-y\vert\rightarrow\infty$. We use a 
unit ultraviolet cutoff, built in as a smooth momentum space 
regulator, whereby $C_{\infty}(x,y)$ becomes a regular function 
of $\vert x-y\vert$ at short distances. 

The following renormalization group \cite{Gallavotti:1985} is 
custom built to study the properties of \REF{0.1}. 
Let $D_{L}$ be the dilatation operator 
given by $D_{L}^{\MRM{T}}(f)(x)=L^{1-\frac{D+\epsilon}{2}}\;
f\left(\frac{x}{L}\right)$. We decompose $C_{\infty}$ into 
$C_{\infty}=D_{L}^{\MRM{T}}\,C_{\infty}\,D_{L}+C_{L}$ such 
that $C_{L}(x,y)$ is of the order 
$O\left(F\left(\frac{\vert x-y\vert}{L}\right)\right)$ 
at $\vert x-y\vert\rightarrow\infty$ with a
rapidly decreasing function $F(t)$. (We will choose an 
exponential cutoff where $F(t)=\MRM{e}^{-\frac{t^{2}}{4}}$.)
For $C_{L}(x,y)$ to be short ranged, the asymptotic of 
$C_{\infty}(x,y)$ and its scaled version
$(D_{L}^{\MRM{T}}\,C_{\infty}\,D_{L})
(x,y)=L^{2-D-\epsilon}\;C\left(\frac{x}{L},\frac{y}{L}\right)$
have to agree. This condition determines the scaling dimension in 
the dilatation operator $D_{L}$. We cannot 
decompose $C_{\infty}$ with an anomalous dilatation, unless 
we give up that $C_{L}$ be short ranged. At this instant, we
have a pair of operators $(C_{L},D_{L})$, where $C_{L}$ is a 
short ranged covariance, and where $D_{L}$ is a dilatation. 
With this pair $(C_{L},D_{L})$ is associated a renormalization 
group transformation
\begin{equation} 
R(C_{L},D_{L})(Z)(\Phi)=
   \frac{\int\MRM{d}\mu_{C_{L}}(\phi)\;
   Z\bigl(D_{L}^{\MRM{T}}(\Phi)+\phi\bigr)}
   {\int\MRM{d}\mu_{C_{L}}(\phi)\;Z(\phi)}
\label{0.2}
\end{equation}
of the perturbation $Z(\phi)$. It defines a representation of
the semi-group of scale transformations with scale factor 
$L>1$ on a suitable space of perturbations $Z(\phi)$. 
(For $L<1$, $C_{L}$ becomes negative.) Transformations of this 
kind have a wide range of applications. See 
\cite{Benfatto/Gallavotti:1995} for a recent review. 

We will show that the transformation given by $(C_{L},D_{L})$ is 
equivalent to the transformation given by another pair $(c_{L},d_{L})$. 
The equivalence is of the form
\begin{equation}
\begin{CD}
Z(\phi)@<{Q(K_{\star})}<<z(\phi)\\
@V{R(C_{L},D_{L})}VV @VV{R(c_{L},d_{L})}V\\
Z(\phi\vert L)@<<{Q(K_{\star})}<z(\phi\vert L)
\end{CD}
\label{0.3}
\end{equation}
where $Q(K_{\star})(z)(\phi)=Z_{\star}(\phi)\;z(\phi)$ and 
$Z_{\star}(\phi)=\MRM{e}^{-\frac{1}{2}(\phi,K_{\star}^{-1}\phi)}$ 
is a short ranged Gaussian fixed point of $R_{L}(C_{L},D_{L})$. The new 
pair $(c_{L},d_{L})$ defines a scale decomposition $c_{\infty}=
d_{L}^{\MRM{T}}\,c_{\infty}\,d_{L}+c_{L}$ of a covariance 
$c_{\infty}$.
The two formulations will be called the massless picture and the 
massive picture respectively. They are well known in the hierarchical 
(or ultra-local)
approximation \cite{Koch/Wittwer:1991}. In the massive picture, both 
$c_{L}$ and $c_{\infty}$ are short ranged. Therefore, the scaling 
dimension of $d_{L}$ is variable. In the massive picture, we will 
show the transformation defined by yet another pair $(\gamma_{L},
\delta_{L})$, which defines a global (space-time independent) scale 
decomposition of $c_{\infty}$, with $\delta_{L}$ the multiplication 
by a global factor $L^{-1}$, is a symmetry. Renormalization group
and symmetry together spread a perturbation $z(\phi)$ to a surface 
\begin{equation}
\begin{CD}
z(\phi)@>{R(\gamma_{L^{\prime}},\delta_{L^{\prime}})}>>
   z(\phi\vert 1,L^{\prime})\\
@V{R(c_{L},d_{L})}VV @VV{R(c_{L},d_{L})}V\\
z(\phi\vert L,1)@>>{R(\gamma_{L^{\prime}},\delta_{L^{\prime}})}>
   z(\phi\vert L,L^{\prime})
\end{CD}
\label{0.4}
\end{equation}
We will then use the equivalence to translate the symmetry back to 
the massless picture. The composition of a renormalization group 
transformation, which scales by a factor $L$, and a symmetry 
transformation, which scales by a factor $L^{\frac{\eta}{2}}$, defines 
an anomalous scale decomposition of $c_{\infty}$. 

Inspired by the construction of a non-trivial $\phi^{4}_{4,\epsilon}$ 
fixed point in the planar approximation \cite{Felder:1985} and beyond 
it \cite{Brydges/Dimock/Hurd:1997}, we consider the renormalization 
group given by $C_{\infty}=F(-\bigtriangleup,\epsilon)\;
(-\bigtriangleup)^{-1+\frac{\epsilon}{2}}$, where $\epsilon$ is 
a small parameter, and where $F(-\bigtriangleup,\epsilon)$ is a 
certain cutoff function. We will argue that the two cases $\epsilon=0$ and 
$\epsilon\neq 0$ are different. We will briefly discuss the relation
between the anomalous dimension $\eta$ and the symmetry applying 
the concepts of \cite{Bell/Wilson:1975} to this case. 
The anomalous dimension of the $\phi^{4}_{4,\epsilon}$ fixed point is 
by definition zero. 

\section{Renormalization group -- massless picture}

We consider renormalization groups of transformations $R(C_{L},D_{L})$,
defined by a pair of operators $(C_{L},D_{L})$, where $C_{L}$ is a 
covariance, $D_{L}$ is a dilatation, and $L$ is a scale, with $L>1$. 
Our starting point and setup is a particular pair $(C_{L},D_{L})$, which 
defines the renormalization group in the massless picture.

\subsection{Covariance $C_{L}$}

Let $C_{L}$ be the covariance operator 
$C_{L}(f)(x)=\int\MRM{d}^{D}y\;C_{L}(x,y)\;f(y)$ given by
\begin{align}
C_{L}(x,y)&=
   \int\frac{\MRM{d}^{D}p}{(2\pi)^{D}}\;\MRM{e}^{\MRM{i}\,p\,(x-y)}\;
   \widehat{C_{L}}(p),
\label{1.1}\\
\widehat{C_{L}}(p)&=
   \frac{\widehat{\chi}(p)-\widehat{\chi}(L\,p)}
   {(p^{2})^{1-\frac{\epsilon}{2}}},
\label{1.2}
\end{align}
where $\widehat{\chi}(p)$ is a cutoff function, and $\epsilon$ is a 
small parameter (possibly zero). 
Let $\widehat{\chi}(p)=F(p^{2})$, where $F(s)$ is a 
monotonously decreasing function, which takes values between $F(0)=1$ 
and $\lim_{s\rightarrow\infty}F(s)=0$. We choose the incomplete 
$\Gamma$-function 
\begin{equation}
\widehat{\chi}(p)\;=\;
   \frac{
      \int_{p^{2}}^{\infty}\frac{\MRM{d}\alpha}{\alpha}\;
      \alpha^{1-\frac{\epsilon}{2}}\;
      \MRM{e}^{-\alpha}
   }{
      \Gamma\left(1-\frac{\epsilon}{2}\right)
   }
\;=\;
   \frac{\Gamma\left(1-\frac{\epsilon}{2},p^{2}\right)}
   {\Gamma\left(1-\frac{\epsilon}{2}\right)}.
\label{1.3}
\end{equation}
For $\epsilon=0$, it becomes $\widehat{\chi}(p)=\MRM{e}^{-p^{2}}$.
The advantages of this particular cutoff function are apparent from 
the integral representations 
\begin{align}
\widehat{C_{L}}(p)&\;=\;
   \frac{
      \int_{1}^{L^{2}}\frac{\MRM{d}\alpha}{\alpha}\;
      \alpha^{1-\frac{\epsilon}{2}}\;
      \MRM{e}^{-\alpha\,p^{2}}
   }{
      \Gamma\left(1-\frac{\epsilon}{2}\right)
   },
\label{1.4}\\
C_{L}(x,y)&\;=\;
   \frac{
      \int_{L^{-2}}^{1}\frac{\MRM{d}\alpha}{\alpha}\;
      \alpha^{\frac{D+\epsilon}{2}-1}\;
      \MRM{e}^{-\alpha\left(\frac{x-y}{2}\right)^{2}}
   }{
      (4\pi)^{\frac{D}{2}}\;
      \Gamma\left(1-\frac{\epsilon}{2}\right)
   }.
\label{1.5}
\end{align}
The covariance $C_{L}$ is a two sided regularization of 
$(-\bigtriangleup)^{-1+\frac{\epsilon}{2}}$, with unit ultraviolet 
cutoff and infrared cutoff $L^{-1}$ (in units of mass). We included 
a parameter $\epsilon$ for the sake of generality, and to distinguish 
it from another parameter, the anomalous dimension $\eta$ to appear 
below. We will assume that $D+\epsilon >2$.

\subsection{Dilatation $D_{L}$}

When the infrared cutoff is removed, we obtain a massless covariance 
$C_{\infty}=\lim_{L\rightarrow\infty}C_{L}$ with unit ultraviolet cutoff.
In this limit, \REF{1.4} and \REF{1.5} become 
\begin{align}
\widehat{C_{\infty}}(p)&\;=\;
   \frac{
      \Gamma\left(1-\frac{\epsilon}{2},p^{2}\right)
   }{
      \Gamma\left(1-\frac{\epsilon}{2}\right)\;
   (p^{2})^{1-\frac{\epsilon}{2}}
   },
\label{1.6}\\
C_{\infty}(x,y)&\;=\;
   \frac{
      \gamma\left(\frac{D+\epsilon}{2}-1,
      \left(\frac{x-y}{2}\right)^{2}\right)
   }{
      (4\pi)^{\frac{D}{2}}\;
      \Gamma\left(1-\frac{\epsilon}{2}\right)\;
      \left(
         \left(\frac{x-y}{2}\right)^{2}
      \right)^{\frac{D+\epsilon}{2}-1}
   }.
\label{1.7}
\end{align}
In other words, $C_{\infty}=\chi/(-\bigtriangleup)^{1-\frac{\epsilon}{2}}$.
Conversely, $C_{L}$ is the difference of $C_{\infty}$ and a scaled 
version of $C_{\infty}$. We have that
\begin{align}
\widehat{C_{L}}(p)&=
   \widehat{C_{\infty}}(p)-
   L^{2-\epsilon}\;\widehat{C_{\infty}}(L\,p),
\label{1.8}\\
C_{L}(x,y)&=
   C_{\infty}(x,y)-
   L^{2-D-\epsilon}\;C_{\infty}\left(\frac{x}{L},\frac{y}{L}\right).
\label{1.9}
\end{align}
Eqs.\ \REF{1.8} and \REF{1.9} read $C_{L}=C_{\infty}-D_{L}^{\MRM{T}}\,
C_{\infty}\,D_{L}$ in operator notation, where $D_{L}$ is the 
dilatation operator, and where $D_{L}^{\MRM{T}}$ is the transposed
of $D_{L}$, given by
\begin{align}
D_{L}^{\MRM{T}}(f)(x)&=
   L^{1-\frac{D+\epsilon}{2}}\;
   f\left(\frac{x}{L}\right),
\label{1.10}\\
D_{L}(f)(x)&=
   L^{1+\frac{D-\epsilon}{2}}\;
   f(L\,x)
\label{1.11}   
\end{align}
respectively.
Thus $D_{L}$ represents a dilatation by the scale factor $L$ on function
space $\{f:\MBB{R}^{D}\rightarrow\MBB{R}\}$, with scaling dimension 
$\sigma=1+\frac{D-\epsilon}{2}$. The dilatations by scale factors $L>1$ 
form a semi-group.

To summarize, we have a pair of operators $(C_{L},D_{L})$, where 
$C_L$ is a short ranged covariance, $D_L$ is a dilatation, and $L$ is a 
scale such that
$D_{1}=1$, $D_{L}\;D_{L^{\prime}}=D_{L\,L^{\prime}}$, and 
$C_{L}=C_{\infty}-D_{L}^{\MRM{T}}\;C_{\infty}\;D_{L}$.
We say that $(C_{L},D_{L})$ defines a scale decomposition of $C_{\infty}$ 
with respect to $D_{L}$.

\subsection{Transformation $R(C_{L},D_{L})$}

With the pair $(C_{L},D_{L})$ is associated a renormalization group 
transformation $R(C_{L},D_{L})$, defined by
\begin{equation}
R(C_{L},D_{L})(Z)(\Phi)=
   \frac{\int\MRM{d}\mu_{C_{L}}(\phi)\;Z(D_{L}^{\MRM{T}}(\Phi)+\phi)}
   {\int\MRM{d}\mu_{C_{L}}(\phi)\;Z(\phi)},
\label{1.12}
\end{equation}
where $\MRM{d}\mu_{C_{L}}(\phi)$ is the Gaussian measure with mean
zero and covariance $C_{L}$, and $Z(\phi)$ is a perturbation, which 
is normalized such that $Z(0)=1$. Recall that
\begin{equation}
\int\MRM{d}\mu_{C_{L}}(\phi)\;\MRM{e}^{(\phi,f)}\;=\;
   \MRM{e}^{\frac{1}{2}(f,C_{L}\,f)}.
\label{1.12a}
\end{equation}

\subsubsection{Averaged perturbation $\MCA{Z}(\Phi)$}

Admit a brief detour around the physical meaning of \REF{1.12}.
We consider effective quantum field theories defined by the covariance 
$C_{\infty}$ and an interaction $V(\phi)=-\log Z(\phi)$. Their physical 
content is coded in the generating function
\begin{equation}
\MCA{G}(f)\;=\;
   \frac{\int\MRM{d}\mu_{C_{\infty}}(\phi)\;
   Z(\phi)\;\MRM{e}^{(\phi,f)}}
   {\int\MRM{d}\mu_{C_{\infty}}(\phi)\;Z(\phi)}
\;=\;
   \MRM{e}^{-\frac{1}{2}\,(f,C_{\infty}\,f)}\;
   \MCA{Z}\bigl(C_{\infty}(f)\bigr),
\label{1.13}
\end{equation}
where $\MCA{Z}(\Phi)$ is the convolution of $Z(\phi)$ with 
$\MRM{d}\mu_{C_{\infty}}(\phi)$, divided by a normalization constant. 
Its explicit form is
\begin{equation}
\MCA{Z}(\Phi)=
   \frac{\int\MRM{d}\mu_{C_{\infty}}(\phi)\;Z(\Phi+\phi)}
   {\int\MRM{d}\mu_{C_{\infty}}(\phi)\;Z(\phi)}.
\label{1.14}
\end{equation}
The transformation $R(C_{L},D_{L})$ performs a dilatation of 
$\MCA{Z}(\Phi)$. We have that
\begin{equation}
\MCA{Z}(D_{L}^{\MRM{T}}(\Phi))=
   \frac{\int\MRM{d}\mu_{C_{\infty}}(\phi)\;
   R(C_{L},D_{L})(Z)(\Phi+\phi)}
   {\int\MRM{d}\mu_{C_{\infty}}(\phi)\;
   R(C_{L},D_{L})(Z)(\phi)}.
\label{1.15}
\end{equation}
The transformation $R(C_{L},D_{L})$ thus enables a study of the behavior
of $\MCA{Z}(\Phi)$ under $D_{L}^{\MRM{T}}$, without having to perform 
the convolution with respect to $\MRM{d}\mu_{C_{\infty}}(\phi)$, which
is difficult because $C_{\infty}$ is long range.

If $R(C_{L},D_{L})(Z_{\star})(\Phi)=Z_{\star}(\Phi)$ then 
$\MCA{Z}\bigl(D_{L}^{\MRM{T}}(\Phi)\bigr)=\MCA{Z}(\Phi)$. In other
words, if we have a fixed point of the renormalization group, then
we have dilatation invariant correlators.

\subsubsection{Semi-group property of $R(C_{L},D_{L})$}

We are lead to consider the following structure. We have a pair of 
operators $(A,B)$ (with certain properties). With this pair, we associate 
a functional transformation
\begin{equation}
R(A,B)(Z)(\Phi)=
   \frac{\int\MRM{d}\mu_{A}(\phi)\;Z(B^{\MRM{T}}(\Phi)+\phi)}
   {\int\MRM{d}\mu_{A}(\phi)\;Z(\phi)}.
\label{1.16}
\end{equation}
The particular pair $(A,B)=(0,1)$ gives the identity transformation. 
Since $(C_{1},D_{1})=(0,1)$, we have that $R(C_{1},D_{1})=1$.
The composition of two transformation \REF{1.16} is another 
transformation of this kind. Let $(A,B)$ and $(A^{\prime},B^{\prime})$ 
be two pairs of operators. An standard computation of Gaussian 
integrals yields
\begin{equation}
R(A,B)\; R(A^{\prime},B^{\prime})=
   R(B^{\prime\MRM{T}}\,A\,B+A^{\prime},B\,B^{\prime}).
\label{1.17}
\end{equation}
Eq.~\REF{1.17} can be thought of as a product of $(A,B)$ and 
$(A^{\prime},B^{\prime})$, given $(A,B)(A^{\prime},B^{\prime})=
(B^{\prime\MRM{T}}\,A\,B+A^{\prime},B\,B^{\prime})$.
From the two properties $D_{L}\,D_{L^{\prime}}=D_{L\,L^{\prime}}$ and 
$C_{L}=C_{\infty}-D_{L}^{\MRM{T}}\,C_{\infty}\,D_{L}$, it 
follows that $(C_{L},D_{L})\;(C_{L^{\prime}},D_{L^{\prime}})=
(C_{L\,L^{\prime}},D_{L\,L^{\prime}})$. Therefore, the transformation 
$R(C_{L},D_{L})$ satisfies the semi-group property
\begin{equation}
R(C_{L},D_{L})\; R(C_{L^{\prime}},D_{L^{\prime}})=
   R(C_{L\,L^{\prime}},D_{L\,L^{\prime}}).
\label{1.18}
\end{equation}

\subsection{Generator $\dot{R}(\dot{C},\dot{D})$}

The transformation $R(C_{L},D_{L})$ associates with a given 
perturbation $Z(\phi)$ a renormalization group trajectory
\begin{equation}
Z(\phi\vert L)=R(C_{L},D_{L})(Z)(\phi).
\label{1.19}
\end{equation}
Due to the semi-group property, $Z(\phi\vert L)$ is the solution of 
the discrete flow equation
\begin{equation}
R(C_{L},D_{L})(Z)(\phi\vert L^{\prime})=Z(\phi\vert L\,L^{\prime})
\label{1.20}
\end{equation}
to the initial condition $Z(\phi\vert 1)=Z(\phi)$. The discrete flow 
equation implies a continuous flow equation for $Z(\phi\vert L)$,
the renormalization group differential equation
\begin{equation}
L\frac{\partial}{\partial L}Z(\phi\vert L)=
   \dot{R}(\dot{C},\dot{D})(Z)(\phi\vert L),
\label{1.21}
\end{equation}
where $\dot{R}(\dot{C},\dot{D})=\frac{\partial}{\partial L}
\big\vert_{L=1}R(C_{L},D_{L})$ is the generator of the renormalization 
group. Its explicit form is
\begin{align}
L\frac{\partial}{\partial L}Z(\phi\vert L)&=
   \left[
      \frac{1}{2}\left(\frac{\delta}{\delta\phi},
      \dot{C}\frac{\delta}{\delta\phi}\right)+
      \left(\dot{D^{\MRM{T}}}\phi,\frac{\delta}{\delta\phi}\right)
   \right]
   Z(\phi\vert L)
\nonumber\\&\phantom{=}-
   Z(\phi\vert L)
   \left[
      \frac{1}{2}\left(\frac{\delta}{\delta\phi},
      \dot{C}\frac{\delta}{\delta\phi}\right)+
      \left(\dot{D^{\MRM{T}}}\phi,\frac{\delta}{\delta\phi}\right)
   \right]_{\phi=0}
   Z(\phi\vert L),
\label{1.22}
\end{align}
where $\dot{C}=\frac{\partial}{\partial L}\big\vert_{L=1}C_{L}$ 
and $\dot{D^{T}}=\frac{\partial}{\partial L}\big\vert_{L=1}
D_{L}^{\MRM{T}}$. See \cite{Glimm/Jaffe:1987,Wilson/Kogut:1974}. 
For the interaction $V(\phi\vert L)=-\log
Z(\phi\vert L)$, eq.~\REF{1.22} implies that
\begin{align}
L\frac{\partial}{\partial L}V(\phi\vert L)&=
   \left[
      \frac{1}{2}\left(\frac{\delta}{\delta\phi},
      \dot{C}\frac{\delta}{\delta\phi}\right)+
      \left(\dot{D^{\MRM{T}}}\phi,\frac{\delta}{\delta\phi}\right)
   \right]
   V(\phi\vert L)
\nonumber\\&\phantom{=}-
   \frac{1}{2}\left(\frac{\delta}{\delta\phi}V(\phi\vert L),
   \dot{C}\;\frac{\delta}{\delta\phi}V(\phi\vert L)\right)
\nonumber\\&\phantom{=}-
   E(\dot{C})(V(\cdot\vert L)),
\label{1.23}
\end{align}
where $E(\dot{C})(V(\cdot\vert L))$ denotes the (field independent)
normalization constant
\begin{align}
E(\dot{C})(V(\cdot\vert L))&=
   \frac{1}{2}\;
   \biggl[
      \left(
         \frac{\delta}{\delta\phi},\dot{C}\;
         \frac{\delta}{\delta\phi}      
      \right)V(\phi\vert L)
\nonumber\\&\phantom{=}-   
      \left(
          \frac{\delta}{\delta\phi}V(\phi\vert L),\dot{C}\;
          \frac{\delta}{\delta\phi}V(\phi\vert L)
      \right)   
   \biggr]_{\phi=0}.
\label{1.24}
\end{align}
Indispensable intermediate volume cutoffs are here treated casually.
For the instant pair $(C_{L},D_{L})$, the differential version 
$(\dot{C},\dot{D})$ is explicitely given by
\begin{align}
\widehat{\dot{C}}(p)&=
   \frac{
      2\;\MRM{e}^{-p^{2}}
   }{
     \Gamma\left(1-\frac{\epsilon}{2}\right)
   },
\label{1.25}\\
\dot{C}(x,y)&=
   \frac{
      2\;
     \MRM{e}^{-\left(\frac{x-y}{2}\right)^{2}}
   }{
      (4\pi)^{\frac{D}{2}}\;
      \Gamma\left(1-\frac{\epsilon}{2}\right)
   },
\label{1.26}
\end{align}
together with
\begin{align}
\dot{D^{\MRM{T}}}(f)(x)&=
   \left[1-\frac{D+\epsilon}{2}-
   x\frac{\partial}{\partial x}\right]\;f(x),
\label{1.27}\\
\widehat{\dot{D^{\MRM{T}}}(f)}(p)&=
   \left[1+\frac{D-\epsilon}{2}+
   p\frac{\partial}{\partial p}\right]\;\widehat{f}(p).
\label{1.28}
\end{align}
The differential pair $(\dot{C},\dot{D})$ in the flow 
equations \REF{1.22} and \REF{1.23} is independent of 
$L$. This is a characteristic feature of the
renormalization group with rescaling. 

To summarize, the renormalization group associated with the 
pair of operators $(C_{L},D_{L})$ comes in two equivalent formulations. 
In its discrete version, one deals with difference equations.
In its continuous version, one deals with differential 
equations. 

\section{Renormalization group -- massive picture}

In this section, we show that the renormalization group defined by
the pair $(C_{L},D_{L})$ is equivalent to a renormalization group 
defined by another pair $(c_{L},d_{L})$.

\subsection{Gaussian fixed point}

The transformation $R(C_{L},D_{L})$ has a fixed point 
$Z_{\star}(\phi)= \MRM{e}^{-\frac{1}{2}(\phi,K_{\star}^{-1}\,
\phi)}$, with $R(C_{L},D_{L})(Z_{\star})(\phi)\,=\,Z_{\star}(\phi)$ 
for all $L>1$ , given by 
$K_{\star}=
(1-\chi)/(-\bigtriangleup)^{1-\frac{\epsilon}{2}}$. 
With the cutoff function \REF{1.3}, $K_{\star}$ has the integral 
representations
\begin{align}
\widehat{K_{\star}}(p)&=
   \frac{
      \int_{0}^{1}\frac{\MRM{d}\alpha}{\alpha}\;
      \alpha^{1-\frac{\epsilon}{2}}\;
      \MRM{e}^{-\alpha\,p^{2}}
   }{
      \Gamma\left(1-\frac{\epsilon}{2}\right)
   },
\label{2.1}\\
K_{\star}(x,y)&=
   \frac{
      \int_{1}^{\infty}\frac{\MRM{d}\alpha}{\alpha}\;
      \alpha^{\frac{D+\epsilon}{2}-1}\;
      \MRM{e}^{-\alpha\left(\frac{x-y}{2}\right)^{2}}
   }{
      (4\pi)^{\frac{D}{2}}\;
      \Gamma\left(1-\frac{\epsilon}{2}\right)
   }.
\label{2.2}
\end{align} 
$K_{\star}$ is short ranged, and $\widehat{K_{\star}}(p)\,=\,
1/\Gamma\left(2-\frac{\epsilon}{2}\right)+O(p^{2})$. 
More generally, we have a line of fixed points given by 
$K_{\star}+\MRM{const.}/(-\bigtriangleup)^{1-\frac{\epsilon}{2}}$.
All of them are long ranged, except for the special point with
$\MRM{const.}=0$.

\subsection{Transformation $Q(C)$}

Let $C$ be an operator. Let $Q(C)$ be the functional operator 
given by
\begin{equation}
Q(C)(Z)(\phi)=
   \MRM{e}^{-\frac{1}{2}(\phi,C^{-1}\,\phi)}\;
   Z(\phi).
\label{2.3}
\end{equation}
Thus $Q(C)$ multiplies a perturbation with a Gauss function.
The two transformations $R(A,B)$ and $Q(C)$ conspire to the
identity
\begin{equation}
R(A,B)\;Q(C)= 
   Q(B^{-1\MRM{T}}\,\MCA{L}^{-1}\,C\,B^{-1})\;
   R(\MCA{L}\,A,B\,\MCA{L}),
\label{2.4}
\end{equation}
where $\MCA{L}^{-1}=1+A\,C^{-1}$, assuming that $A=A^{\MRM{T}}$, 
$C=C^{\MRM{T}}$, invertibility of $A$, $B$, and $C$, and existence 
of the Gaussian convolutions.

\subsection{Substitution $Z(\phi)=Z_{\star}(\phi)\;z(\phi)$}

We write the perturbation $Z(\phi)$ as a product of the fixed 
point $Z_{\star}(\phi)$ and a new perturbation $z(\phi)$.
Substitute $Z(\phi)=Z_{\star}(\phi)\;z(\phi)$ to obtain 
\begin{equation}
R(C_{L},D_{L})(Z_{\star}\;z)(\phi)=
   Z_{\star}(\phi)\;
   R(c_{L},d_{L})(z)(\phi),
\label{2.5}
\end{equation}
with a new pair $(c_{L},d_{L})$ of operators, defining the massive 
picture of the renormalization group. They are given by 
\begin{equation}
c_{L}=\MCA{L}_{L}\,C_{L},\quad 
d_{L}=D_{L}\,\MCA{L}_{L}, 
\label{2.6}
\end{equation}
where $\MCA{L}_L$ is the operator $\MCA{L}_{L}=
K_{\star}/(K_{\star}+C_{L})$. Notice that $C_{L}$ and 
$K_{\star}$ commute. $\MCA{L}_L$ is explicitely given by
\begin{equation}
\widehat{\MCA{L}_{L}}(p)\;=\;
   \frac{1-\widehat{\chi}(p)}{1-\widehat{\chi}(L\,p)}
\;=\;
   L^{-2+\epsilon}\;\frac{\widehat{K_{\star}}(p)}
   {\widehat{K_{\star}}(L\,p)}.
\label{2.7}
\end{equation}
Think of $\widehat{\MCA{L}_L}(p)$ as a $p$-dependent correction to 
the scale factor in the field dilatation. At $p^{2}=0$ it yields an 
extra factor $L^{-2+\epsilon}$ to the dilataton, while it becomes 
one as $p^{2}\rightarrow\infty$.

\subsection{Covariance $c_{L}$}

Like $C_{L}$, the  covariance $c_{L}$ is short ranged. The explicit form 
of $c_{L}$ is 
\begin{equation}
\widehat{c_{L}}(p)\;=\;
   \widehat{\MCA{L}_L}(p)\;\widehat{C_L}(p)\;=\;
   \frac{1-\widehat{\chi}(p)}{1-\widehat{\chi}(L\,p)}\;
   \frac{\widehat{\chi}(p)-\widehat{\chi}(L\,p)}
   {(p^{2})^{1-\frac{\epsilon}{2}}}.
\label{2.8}
\end{equation}
Analogous to $C_{1}=0$, it satisfies $c_{1}=0$. In the limit
$L\rightarrow\infty$, \REF{2.8} turns into a covariance
\begin{equation}
\widehat{c_{\infty}}(p)\;=\;
   \frac{\bigl(1-\widehat{\chi}(p)\bigr)\;
   \widehat{\chi}(p)}{(p^{2})^{1-\frac{\epsilon}{2}}}
\;=\;
   \bigl(1-\widehat{\chi}(p)\bigr)\;
   \widehat{C_{\infty}}(p).
\label{2.9}
\end{equation}
Recall that $C_{\infty}$ is a massless covariance with unit ultraviolet
cutoff. In contrast, $c_{\infty}$ is a massive (short ranged) covariance 
with twosided unit cutoffs.

\subsection{Dilatation $d_{L}$}

From \REF{2.7}, it follows that $\MCA{L}_1=1$ and consequently 
$d_{1}=1$, in analogy to $D_{1}=1$. Furthermore, $d_{L}$ satisfies 
the composition law $d_{L}\;d_{L^{\prime}}=d_{L\,L^{\prime}}$. 
The pair $(c_{L},d_{L})$ can be shown to be a scale decomposition of the 
covariance $c_{\infty}$ with respect to $d_{L}$. We have that
$c_{L}=c_{\infty}-d_{L}^{\MRM{T}}\,c_{\infty}\,d_{L}$. Therefrom, 
we conclude that $R(c_{L},d_{L})$ satisfies the semi-group 
properties $R(c_{1},d_{1})=1$ and
\begin{equation}
R(c_{L},d_{L})\;R(c_{L^{\prime}},d_{L^{\prime}})\;=\;
   R(c_{L\,L^{\prime}},d_{L\,L^{\prime}}).
\label{2.10}
\end{equation}
The reasoning is completely analogous to the massless case, since 
the algebraic properties of the pair $(c_{L},d_{L})$ are identical 
to those of $(C_{L},D_{L})$. 

\subsection{Generator $\dot{R}(\dot{c},\dot{d})$}

To compute the renormalization group differential
equation in the massive picture, all we have to do, is to compute the 
differential pair $(\dot{c},\dot{d})$ and to replace $(\dot{C},\dot{D})$ 
by it in \REF{1.22} and \REF{1.23}. The differential covariance and 
dilatation pick up an extra term $\dot{\MCA{L}}$ as compared to the 
massless picture. The extra term is
\begin{equation}
\widehat{\dot{\MCA{L}}}(p)\;=\;
   \frac{\partial}{\partial L}\bigg\vert_{L=1}
   \widehat{\MCA{L}_{L}}(p)
\;=\;
   \frac{p\nabla_{p}\widehat{\chi}(p)}
   {1-\widehat{\chi}(p)}.
\label{2.11}
\end{equation}
With our cutoff function \REF{1.3}, the extra term becomes
\begin{equation}
\widehat{\dot{\MCA{L}}}(p)\;=\;
   \frac{
      -2\,(p^{2})^{1-\frac{\epsilon}{2}}\;
      \MRM{e}^{-p^{2}}\;   
   }{
      \gamma \left(1-\frac{\epsilon}{2},p^{2}\right)
   }
\;=\;
   \frac{-2}
   {\int_{0}^{1}\MRM{d}\alpha\;(1-\alpha)^{\frac{-\epsilon}{2}}\;
   \MRM{e}^{\alpha\,p^{2}}}.
\label{2.12}
\end{equation}
In the special case when $\epsilon=0$, it takes the form of the 
generating function of the Bernoulli numbers, namely
$\widehat{\dot{\MCA{L}}}(p)=-2\,p^{2}/(\MRM{e}^{p^{2}}-1)$.
In the massive picture, we thus have a differential pair of 
operators $(\dot{c},\dot{d})$, given by
\begin{equation}
\widehat{\dot{c}}(p)\;=\;
   \widehat{\dot{\MCA{L}}}(p)+\widehat{\dot{C}}(p)
\;=\;
   2\,\MRM{e}^{-p^{2}}\;
   \left(
      \frac{1}{\Gamma\left(1-\frac{\epsilon}{2}\right)}-
      \frac{(p^{2})^{1-\frac{\epsilon}{2}}}
      {\gamma\left(1-\frac{\epsilon}{2},p^{2}\right)}
   \right)
\label{2.13}
\end{equation}
together with
\begin{equation}
\widehat{\dot{d^{\MRM{T}}}(f)}(p)\;=\;
   \widehat{\dot{\MCA{L}}(f)}(p)+
   \widehat{\dot{D^{\MRM{T}}}(f)}(p)
\;=\;
   \left[
      \widehat{\dot{\MCA{L}}}(p)+
      1+\frac{D-\epsilon}{2}+
      p\frac{\partial}{\partial p}
   \right]\;\widehat{f}(p).
\label{2.14}
\end{equation}
The change from the massless picture to the massive picture 
is thus particularly simple in the differential formulation 
of the renormalization group. Finally, we remark that
$d_{L}=D_{L}\,\MCA{L}_{L}\,D_{L}^{-1}\,D_{L}$ with
\begin{equation}
\widehat{(D_{L}\MCA{L}_{L}D_{L}^{-1})}(p)\;=\;
   \widehat{\MCA{L}_{L}}(L^{-1}p)
\;=\;
   \frac{1-\widehat{\chi}(L^{-1}\,p)}
   {1-\widehat{\chi}(p)}.
\label{2.15}
\end{equation}
Which of the two pictures is better suited, depends on the kind of 
perturbation one is considering, whether it is functionally closer 
to the trivial fixed point or to the Gaussian fixed point. In the 
hierarchical approximation for instance, the massive picture has proved 
to be better suited for an investigation of the non-trivial scalar 
fixed point in three dimensions \cite{Koch/Wittwer:1991}. 

\section{Symmetry -- massive picture}

In this section, we show that the renormalization group defined by
the pair $(c_{L},d_{L})$ has a symmetry, the transformation defined
by another pair $(\gamma_{L},\delta_{L})$.

\subsection{Symmetry pair $(\gamma_{L},\delta_{L})$}

Let $L$ be a scale, with $L>1$. Let $\gamma_{L}$ and $\delta_{L}$
be the operators given by 
\begin{equation}
\gamma_{L}=\left(1-L^{-2}\right)\;c_{\infty},\quad
\delta_{L}=L^{-1}. 
\label{3.1}
\end{equation}
The operator $\delta_{L}$ 
performs a global (space-time independent) multiplication by 
$L^{-1}$. Obviously, $(\gamma_{L},\delta_{L})$ is a (global) scale 
decomposition of $c_{\infty}$ with respect to $\delta_{L}$,
namely $\gamma_{L}=c_{\infty}-\delta_{L}^{\MRM{T}}\;c_{\infty}\;
\delta_{L}$. It follows that the transformation $R(\gamma_{L},
\delta_{L})$ satisfies the semi-group properties $R(\gamma_{1},
\delta_{1})=1$ and
\begin{equation}
R(\gamma_{L},\delta_{L})\;R(\gamma_{L^{\prime}},\delta_{L^{\prime}})\;=\;
   R(\gamma_{L\,L^{\prime}},\delta_{L\,L^{\prime}}).
\label{3.2}
\end{equation}

\subsection{Commutator of $R(c_{L},d_{L})$ and $R(\gamma_{L^{\prime}},
\delta_{L^{\prime}})$}

The two one parameter families of transformations $R(c_{L},d_{L})$ and 
$R(\gamma_{L^{\prime}},\delta_{L^{\prime}})$ commute: For all $L>1$ and 
$L^{\prime}>1$, we have that
\begin{equation}
R(c_{L},d_{L})\;R(\gamma_{L^{\prime}},\delta_{L^{\prime}})=
   R(\gamma_{L^{\prime}},\delta_{L^{\prime}})\;R(c_{L},d_{L}).
\label{3.3}
\end{equation} 
This property defines a symmetry of $R(c_{L},d_{L})$. An analogous 
construction can be made in the massless picture. The point here is
that the symmetry covariance is short ranged in the massive picture.
To prove \REF{3.3}, we compute
\begin{equation}
(c_{L},d_{L})\;(\gamma_{L^{\prime}},\delta_{L^{\prime}})=
   (\delta_{L^{\prime}}^{\MRM{T}}\,c_{L}\,\delta_{L^{\prime}}+
   \gamma_{L^{\prime}},d_{L}\,\delta_{L^{\prime}})
\label{3.4}
\end{equation}
and compare the result with
\begin{equation}
(\gamma_{L^{\prime}},\delta_{L^{\prime}})\;(c_{L},d_{L})=
   (d_{L}^{\MRM{T}}\,\gamma_{L^{\prime}}\,d_{L}+c_{L},
   \delta_{L^{\prime}}\,d_{L}).
\label{3.5}
\end{equation}
The two dilatations commute, $d_{L}\,\delta_{L^{\prime}}=
\delta_{L^{\prime}}\,d_{L}$. The combined covariances are
\begin{align}
\delta_{L^{\prime}}^{\MRM{T}}\,c_{L}\,\delta_{L^{\prime}}+
\gamma_{L^{\prime}}&=
   (L^{\prime})^{-2}\;
   \biggl(
      c_{\infty}-d_{L}^{\MRM{T}}\,c_{\infty}\,d_{L}
   \biggr)+
   \biggl(
      1-(L^{\prime})^{-2}
   \biggr)
   c_{\infty}
\nonumber\\&=
   c_{\infty}-
   (L^{\prime})^{-2}\;
   d_{L}^{\MRM{T}}\,c_{\infty}\,d_{L}
\label{3.6}
\end{align}
and
\begin{align}
d_{L}^{\MRM{T}}\,\gamma_{L^{\prime}}\,d_{L}+c_{L}&=
   \biggl(
      1-(L^{\prime})^{-2}
   \biggr)\;
   d_{L}^{\MRM{T}}\,c_{\infty}\,d_{L}+
   c_{\infty}-
   d_{L}^{\MRM{T}}\,c_{\infty}\,d_{L}
\nonumber\\&=
   c_{\infty}-
   (L^{\prime})^{-2}\;
   d_{L}^{\MRM{T}}\,c_{\infty}\,d_{L}.
\label{3.7}
\end{align}
These two covariances are the same. It follows that $R(c_{L},d_{L})$ and 
$R(\gamma_{L^{\prime}},\delta_{L^{\prime}})$ indeed commute.

\subsection{Composed transformation}

To obtain a better understanding of the symmetry transformation, we 
can study the composition of a renormalization group transformation
and a symmetry transformation. This composition turns out to be 
given by yet another pair. Let $L^{\prime}=L^{\frac{\eta}{2}}$,
where $\eta$ parametrizes the symmetry transformation. (A special 
value of $\eta$ will be identified with the anomalous dimension.) Then 
we have that
\begin{equation}
R(c_{L},d_{L})\,R(\gamma_{L^{\frac{\eta}{2}}},
\delta_{L^{\frac{\eta}{2}}})=
R(c_{L,\eta},d_{L,\eta}),
\label{3.8}
\end{equation}
where
\begin{equation}
d_{L,\eta}\;=\;
   d_{L}\,\delta_{L^{\frac{\eta}{2}}}
\;=\;
   L^{-\frac{\eta}{2}}\;
   d_{L},
\label{3.9}
\end{equation}
and
\begin{equation}
c_{L,\eta}\;=\;
   c_{\infty}-
   L^{-\eta}\;
   d_{L}^{\MRM{T}}\,c_{\infty}\,d_{L}
\;=\;
   c_{\infty}-
   d_{L,\eta}^{\MRM{T}}\,c_{\infty}\,d_{L,\eta}.
\label{3.10}
\end{equation}
In other words, the composition of the renormalization group and 
the symmetry is a renormalization group built from an anomalous 
scale decomposition $(c_{L,\eta},d_{L,\eta})$ of $c_{\infty}$.

\subsection{Differential pair $(\dot{\gamma},\dot{\delta})$}

The generator of the symmetry is particularly simple. It is 
given by the differential pair of operators $(\dot{\gamma},
\dot{\delta})$ with $\dot{\gamma}=2\,c_{\infty}$ and $\dot{\delta}=
-1$. The generators of the renormalization group and the symmetry are
\begin{align}
r&=
   \frac{1}{2}\;
   \left(\frac{\delta}{\delta\phi},\dot{c}
   \frac{\delta}{\delta\phi}\right)+
   \left(\dot{d^{\MRM{T}}}\phi,\frac{\delta}{\delta\phi}\right),
\label{3.11}\\
u&=
   \frac{1}{2}\;
   \left(\frac{\delta}{\delta\phi},\dot{\gamma}
   \frac{\delta}{\delta\phi}\right)+
   \left(\dot{\delta^{\MRM{T}}}\phi,\frac{\delta}{\delta\phi}\right).
\label{3.12}
\end{align}
They satisfy $[r,u]=0$, as can be checked directly. Renormalization 
group and symmetry associate with a perturbation $z(\phi)$ a two 
parametric orbit
\begin{equation}
z(\phi\vert L,L^{\prime})\;=\;
R(c_{L},d_{L})\,R(\gamma_{L^{\prime}},\delta_{L^{\prime}})(z)(\phi),
\label{3.13}
\end{equation}
whose dependence on $L$ is governed by \REF{3.11}, and whose 
dependence on $L^{\prime}$ is governed by \REF{3.12}. Indeed, 
\REF{3.13} satisfies the differential equations
\begin{align}
L\frac{\partial}{\partial L}z(\phi\vert L,L^{\prime})&=
   r(z)(\phi\vert L,L^{\prime})
   -z(\phi\vert L,L^{\prime})\;
   r(z)(0\vert L,L^{\prime}),
\label{3.13a}\\
L^{\prime}\frac{\partial}{\partial L^{\prime}}z(\phi\vert L,L^{\prime})&=
   u(z)(\phi\vert L,L^{\prime})
   -z(\phi\vert L,L^{\prime})\;
   u(z)(0\vert L,L^{\prime}).
\label{3.13b}
\end{align}
The interaction $v(\phi\vert L,L^{\prime})=-\log z(\phi\vert L,L^{\prime})$
obey an exponentiated version \REF{1.27} and \REF{1.28} of \REF{3.13a} and 
\REF{3.13b}.

\section{Symmetry -- massless picture}

In this section, we transform the symmetry $R(\gamma_{L},\delta_{L})$
to the massless picture. Recall that the change of pictures is the 
transformation $Z(\phi)=Q(K_{\star})(z)(\phi)$ and that
\begin{equation}
R(C_{L},D_{L})\;Q(K_{\star})=
   Q(K_{\star})\;R(c_{L},d_{L}).
\label{4.1}
\end{equation}
To translate $R(\gamma_{L},\delta_{L})$ to the massless picture, we
have to conjugate with $Q(K_{\star})$. From \REF{3.8} 
and
\begin{equation}
Q(C)\;Q(D)=
   Q\left(C\frac{1}{C+D}D\right)
\label{4.2}
\end{equation}
it follows that
\begin{equation}
R(A,B)\;Q(C)=
   Q(C)\;Q\left(-C\frac{1}{C^{\prime}-C}C^{\prime}\right)\;
   R(A^{\prime},B^{\prime}),
\label{4.3}
\end{equation}
with $A^{\prime}=\MCA{L}\;A$, $B^{\prime}=B\;\MCA{L}$,
$C^{\prime}=(B^{-1})^{\MRM{T}}\;(C+A)\;B^{-1}$, where
$\MCA{L}=1/(1+A\,C^{-1})$.
In the massless picture, the symmetry transformation therefore consists 
of two parts. One part is the transformation given by a pair of operators.
The second part is the multiplication with a certain Gauss function. 
This term is essentially a kinetic term as we will see.

\subsection{$A^{\prime}$, $B^{\prime}$, $C^{\prime}$, and 
$C^{\prime\prime}$ for the symmetry}

Let $A$, $B$, and $C$ be given by $A=(1-L^{-2})\;c_{\infty}$, 
$B=L^{-1}$, and $C=-K_{\star}$. It follows that $\MCA{L}=
1/(1-(1-L^{-2})\,\chi)$ and thus
\begin{align}
A^{\prime}&=
   \frac{
      (1-L^{-2})\;
      c_{\infty}
   }{
      1-(1-L^{-2})\chi
   },
\label{4.4}\\
B^{\prime}&=
   \frac{L^{-1}}{1-(1-L^{-2})\chi},
\label{4.5}\\
C^{\prime}&=
   \left(L^{2}-1-\frac{L^{2}}{\chi}\right)\;
   c_{\infty},
\label{4.6}
\end{align}
and 
\begin{equation}
\frac{1}{C^{\prime\prime}}\;=\;
   \frac{C-C^{\prime}}{C\,C^{\prime}}
\;=\;
   \frac{
      (1-L^{-2})\;
      (-\bigtriangleup)^{1-\frac{\epsilon}{2}}
   }{
      1-(1-L^{-2})\chi
   }.
\label{4.7}
\end{equation}
The symmetry transformation $R(\gamma_{L},\delta_{L})$ thus has the 
following form in the massless picture. It maps $Z(\phi)$ to a symmetry
orbit
\begin{equation}
Z(\Phi\vert 1,L)=
   \MRM{e}^{-\frac{1}{2}(\Phi,\frac{1}{C^{\prime\prime}}\Phi)}\;
   \frac{\int\MRM{d}\mu_{A^{\prime}}(\phi)\;
   Z((B^{\prime})^{\MRM{T}}(\Phi)+\phi)}
   {\int\MRM{d}\mu_{A^{\prime}}(\phi)\;
   Z(\phi)}.
\label{4.8}
\end{equation}
Remarkably, the function
$\widehat{C^{\prime\prime}}(p)^{-1}=
(1-L^{-2})\;(p^{2})^{1-\frac{\epsilon}{2}}/ 
\bigl(1-(1-L^{-2})\,\widehat{\chi}(p)\bigr)$ 
is an analytic function of $p^{2}$ at the origin only if $\epsilon=0$. 
A straight forward calculation confirms directly that the transformation 
\REF{4.8} satisfies the properties of a semi-group and that it commutes 
with the renormalization group. 

\subsection{$A^{\prime}$, $B^{\prime}$, $C^{\prime}$, and 
$C^{\prime\prime}$ for the composed transformation}

The composition of a symmetry transformation, which scales by a 
factor $L^{\prime}$, and a renormalization group transformation, which
scales by a factor $L$, is computed analogously. Let 
$A=c_{\infty}-(L^{\prime})^{-2}\,d_{L}^{\MRM{T}}\,c_{\infty}\,d_{L}$,
$B=(L^{\prime})^{-1}\,d_{L}$, and $C=-K_{\star}$. Then
\begin{equation}
\MCA{L}=
   \frac{1-\chi_{L}}{1-\chi}\;
   \frac{1}{1-\bigl(1-(L^{\prime})^{-2}\chi_{L}\bigr)},
\label{4.9}
\end{equation}
where $\widehat{\chi_{L}}(p)=\widehat{\chi}(L\,p)$. It follows that
\begin{eqnarray}
A^{\prime}&=&
   \frac{1}{(-\bigtriangleup)^{1-\frac{\epsilon}{2}}}\;
   \frac{\chi\,(1-\chi_{L})
   -(L^{\prime})^{-2}(1-\chi)\,\chi_{L}}
   {1-\bigl(1-(L^{\prime})^{-2}\bigr)\,\chi_{L}},
\label{4.10}\\
B^{\prime}&=&
   D_{L}\;
   \frac{(L^{\prime})^{-1}}
   {1-\bigl(1-(L^{\prime})^{-2}\bigr)\chi_{L}},
\label{4.11}\\
C^{\prime}&=&
   \frac{1}{(-\bigtriangleup)^{1-\frac{\epsilon}{2}}}\;
   (1-\chi)\;
   (L^{\prime})^{2}\;
   \biggl(-1+\bigl(1-(L^{\prime})^{-2}\bigr)\,\chi\biggr),
\label{4.12}
\end{eqnarray}
and
\begin{equation}
\frac{1}{C^{\prime\prime}}=
   \frac{1-(L^{\prime})^{-2}}
   {1-\bigl(1-(L^{\prime})^{-2}\bigr)\chi}\;
   (-\bigtriangleup)^{1-\frac{\epsilon}{2}}.
\label{4.13}
\end{equation}
For $L=1$, one reproduces the formulas \REF{4.4}, \REF{4.5}, and 
\REF{4.6} for the pure symmetry transformation. Notice that 
\REF{4.13} is independent of $L$, as it should. Eq. \REF{4.8}
generalizes to
\begin{equation}
Z(\Phi\vert L,L^{\prime})=
   \MRM{e}^{-\frac{1}{2}(\Phi,\frac{1}{C^{\prime\prime}}\Phi)}\;
   \frac{\int\MRM{d}\mu_{A^{\prime}}(\phi)\;
   Z((B^{\prime})^{\MRM{T}}(\Phi)+\phi)}
   {\int\MRM{d}\mu_{A^{\prime}}(\phi)\;
   Z(\phi)}.
\label{4.14}
\end{equation}

\subsection{Generators in the massless picture}

We compute the generators of the double flow \REF{4.14} for the
interaction $V(\phi\vert L,L^{\prime})=-\log Z(\phi\vert L,L^{\prime})$
from
\begin{eqnarray}
V(\Phi\vert L,L^{\prime})&=&
   \frac{1}{2}\left(\Phi,\frac{1}{C^{\prime\prime}}\Phi\right)
\nonumber\\&\phantom{=}&
   -\log\int\MRM{d}\mu_{A^{\prime}}(\phi)\;
   \MRM{e}^{-V\bigl((B^{\prime})^{\MRM{T}}(\Phi)+\phi\bigr)}
\nonumber\\&\phantom{=}&
   +\log\int\MRM{d}\mu_{A^{\prime}}(\phi)\;
   \MRM{e}^{-V(\phi)},
\label{4.15}
\end{eqnarray} 
where $A^{\prime}$, $B^{\prime}$, and $\frac{1}{C^{\prime\prime}}$
are given by \REF{4.10}, \REF{4.11}, and \REF{4.13} respectively.
We find two flow equations. The first flow equation is identical with 
\REF{1.23} and \REF{1.24}. It describes the flow in the renormalization
group direction. The second flow equation is
\begin{eqnarray}
&&L^{\prime}\frac{\partial}{\partial L^{\prime}}V(\phi\vert L,L^{\prime})=
   \left(\phi,(-\bigtriangleup)^{1-\frac{\epsilon}{2}}\phi\right)
\nonumber\\&\phantom{=}&
   +\left[
      \frac{1}{2}
      \left(
         \frac{\delta}{\delta\phi},
         2\,c_{\infty}\;
         \frac{\delta}{\delta\phi}
      \right)
      +\left(
      (-1+2\chi)\phi,
      \frac{\delta}{\delta\phi}
      \right)
   \right]V(\phi\vert L,L^{\prime})
\nonumber\\&\phantom{=}&
   -\frac{1}{2}
   \left(
      \frac{\delta}{\delta\phi}V(\phi\vert L,L^{\prime}),
      2\,c_{\infty}
      \frac{\delta}{\delta\phi}V(\phi\vert L,L^{\prime})
   \right)
\nonumber\\&\phantom{=}&
   -E(V(\cdot\vert L,L^{\prime})),
\label{4.16}
\end{eqnarray}
where
\begin{eqnarray}
E(V(\cdot\vert L,L^{\prime}))&=&
   \frac{1}{2}
   \biggl[
      \left(
         \frac{\delta}{\delta\phi},
         2\,c_{\infty}
         \frac{\delta}{\delta\phi}V(\phi\vert L,L^{\prime})
      \right)
\nonumber\\&\phantom{=}&
      -\left(
         \frac{\delta}{\delta\phi}V(\phi\vert L,L^{\prime}),
         2\,c_{\infty}
         \frac{\delta}{\delta\phi}V(\phi\vert L,L^{\prime})
      \right)
   \biggr]_{\phi=0}
\label{4.17}
\end{eqnarray}
It describes the flow in the symmetry direction. Both together, plus 
an initial condition, are equivalent to the integral \REF{4.15}.
The novelty as compared to the massive picture is the appearance
of a kinetic term.

\section{Anomalous dimension $\eta$}

This section contains a brief discussion of fixed points with
anomalous dimension based on the following hypothesis,
adapted from \cite{Bell/Wilson:1975} and \cite{Wegner:1976}. 
See also the recent discussion in \cite{Comellas:1997,
Morris/Turner:1997} and references therein.

\subsection{Fixed point $(z_{\star}(\phi),\eta)$}  

We define a fixed point $z(\phi)$ (in the massive picture) with 
anomalous dimension $\eta$ as a fixed point of $R(c_{L},d_{L})\,
R(\gamma_{L^{\frac{\eta}{2}}},\delta_{L^{\frac{\eta}{2}}})$,
the composition of renormalization group transformation 
$R(c_{L},d_{L})$, which scales by a factor $L$, and a symmetry
transformation $R(\gamma_{L^{\frac{\eta}{2}}},
\delta_{L^{\frac{\eta}{2}}})$, which scales by a factor 
$L^{\frac{\eta}{2}}$. 

\subsection{Discussion of $(z_{\star}(\phi),\eta)$}

$z_{\star}(\phi)$ is a fixed point of the renormalization group 
transformation $R(c_{L},d_{L})$ only if $\eta=0$. In this case, 
there are two possibilities. Either $z_{\star}(\phi)$ is also a 
fixed point of $R(\gamma_{L^{\prime}},\delta_{L^{\prime}})$, 
so to speak a double fixed point, or one has a line of fixed points
$z_{\star}(\phi\vert L^{\prime})$ generated as a symmetry orbit 
from an arbitrary representative $z_{\star}(\phi)$. This is the 
case for the trivial fixed point. 

The case with anomalous dimension $\eta$ is a generalization hereof. 
There one has an invariant line of the renormalization group rather 
than a line of fixed points, where the renormalization group acts on
the invariant line by an inverse symmetry transformation. A fixed 
point $z_{\star}(\phi)$ with non-zero anomalous dimension $\eta$ is 
{\sl not} a stationary renormalization group flow. (There remains the 
possibility to declare a composition of the renormalization group and 
the symmetry to be a new renormalization group.) The appropriate 
framework to analyze such a situation is to divide the space of
perturbations into symmetry orbits and to consider the renormalization
group on orbit space. A natural possibility is to consider one 
representative on each orbit by imposing a renormalization condition,
which breaks the symmetry. Such a condition could be to demand 
the pre factor of the kinetic term in the interaction to be a given
number. In the presence of a symmetry, the spectrum of a fixed point
should include a marginal operator, the direction of the symmetry. 

A principal problem is the following. To make the fixed point 
problem tractable, one has to truncate the space of interactions. 
Ideally, the truncation would be such that the symmetry leaves 
invariant the space of truncated interactions. Because then, the 
orbit construction can be done on the truncated space. If however 
the truncation breaks the symmetry, the topic of anomalous dimension
is buried in the no-truncation limit, which is usually very 
difficult to analyze. The idea forwarded in \cite{Bell/Wilson:1975}
for this situation is to truncate in such a way that a marginal 
operator survives truncation. In other words that the truncations
is such that at least differentially the symmetry persists. 
The local approximation to the renormalization group unfortunately
breaks the symmetry, wherefore the concept of anomalous dimension
loses its meaning in the local approximation, and in particular in 
the hierarchical renormalization group.

\subsection{$\epsilon$-model}

Consider the fixed point constructions \cite{Felder:1985} and 
\cite{Brydges/Dimock/Hurd:1997} for the $\epsilon$-model in this 
light. Their constructions work directly in the massless 
picture. When $\epsilon$ is non-zero, the situation is the following.
If one restricts the renormalization group to vertices, which are
$\MCA{C}^{\infty}$-functions of the momenta, then the symmetry can be 
ruled out because it generates a non-$\MCA{C}^{\infty}$ kinetic term.
In this situation, the authors consider an invariant subspace of 
the renormalization group, which is not invariant under the symmetry. 
Remarkably, the authors succeed to construct a non-trivial fixed point 
in the $\MCA{C}^{\infty}$-subspace. (For the planar approximation, 
this is perhaps not so surprising because the planar approximation 
itself also breaks the symmetry.) Such a non-trivial fixed point is 
by definition a fixed point with anomalous dimension zero. 

If on the other hand one wants to investigate flows modulo the symmetry 
(as one has to in the case of non-zero anomalous dimension) then one 
cannot restrict the flow to $\MCA{C}^{\infty}$-vertices
in this model. Whether the $\epsilon$-model has another fixed 
point in this bigger space, I do not know. If this is the case,
it would be very interesting to determine $\eta(\epsilon)$.
In the most interesting case, the three dimensional 
theory with $\epsilon=0$, the situation is different. There the
symmetry preserves the property of analyticity in the momenta. 
Therefore it cannot be ruled out.

\subsection{Differential equation for $v_{\star}(\phi)$}

The fixed point problem \REF{5.1} in the massive picture takes 
the form of the following renormalization group differential equation 
for the fixed point interaction $v_{\star}(\phi)=-\log z_{\star}(\phi)$,
\begin{gather}
   \left[
      \frac{1}{2}\left(\frac{\delta}{\delta\phi},
      [\dot{c}-\eta\; c_{\infty}]\frac{\delta}{\delta\phi}\right)+
      \left(
      \left[\dot{d^{\MRM{T}}}-\frac{\eta}{2}\right]
      \phi,\frac{\delta}{\delta\phi}\right)
   \right]
   v_{\star}(\phi)
\nonumber\\
   -\frac{1}{2}\left(\frac{\delta}{\delta\phi}
   v_{\star}(\phi),
   [\dot{c}-\eta\; c_{\infty}]\;\frac{\delta}{\delta\phi}
   v_{\star}(\phi)\right)
   =E(\dot{c}-\eta\; c_{\infty})(v_{\star}),
\label{5.1}
\end{gather}
where 
\begin{align}
E(\dot{c}-\eta\;c_{\infty})(v_{\star})&=
   \frac{1}{2}\;
   \biggl[
      \left(
         \frac{\delta}{\delta\phi},
         [\dot{c}-\eta\;c_{\infty}]\;
         \frac{\delta}{\delta\phi}      
      \right)v(\phi)
\nonumber\\&\phantom{=}-   
      \left(
          \frac{\delta}{\delta\phi}v_{\star}(\phi),
          [\dot{c}-\eta\; c_{\infty}]\;
          \frac{\delta}{\delta\phi}v_{\star}(\phi)
      \right)   
   \biggr]_{\phi=0}
\label{5.2}
\end{align}
A natural problem is to find good truncation schemes for 
\REF{5.1} and \REF{5.2} and to determine $\eta$ by the 
condition that the spectrum of $v_{\star}(\phi)$ have a 
marginal operator. An investigation of various schemes will 
be presented elsewhere. 

Recall the generators \REF{3.11} and \REF{3.12} of the 
renormalization group and the symmetry respectively. 
If we neglect the trivial normalization constants, the 
fixed point problem becomes a linear eigenvalue problem
\begin{equation}
r(z_{\star})(\phi)\;=\;
-\frac{\eta}{2}\;
u(z_{\star})(\phi)
\label{5.3}
\end{equation}
for $r\,u^{-1}$. The meaning of the anomalous dimension is thus 
to be an eigenvalue. Eq.~\REF{5.1} is just an exponentiated and 
thus non-linear form of \REF{5.3}. The analogous equations in the 
massless picture follow immediately from \REF{4.15} and \REF{4.16}.

\vspace{\baselineskip}\noindent
\begin{minipage}[l]{\textwidth}
{\Large{\bf Acknowledgements}}

\vspace{\baselineskip}
\noindent
I would like to thank Andreas Pordt and Peter Wittwer for 
helpful communications on the anomalous dimension $\eta$, 
and Bernd Gehrmann for computational support.

\end{minipage}

\end{document}